# Upon the Modeling and the Optimization of the Debiting Process through Computer Aided Non-Conventional Technologies

**Tiberiu Marius Karnyanszky [1], Mihai Țîțu [2]**
[1] "Tibiscus" University of Timișoara, Romania
[2] "Lucian Blaga" University of Sibiu, Romania

**ABSTRACT.** The debiting process of the remarkable properties materials can be managed through unconventional technologies as the complex electrical erosion. We present the modeling of the previous experimental results to obtain a mathematical dependence of the output parameters (processing time, surface quality) on the input parameters (voltage or current). All the experimental data are memorized on a database and for each particular debiting process a new dependence is built. Because all the experiments applied in the Romanian laboratories or practical applications of the nonconventional technological processes in the factories were based on the particular conditions of one activity, this papers presents the technical implementation of a computer-aided solution that keeps all previous experimental data, optimizes the processing conditions and eventual manage the driving gear. The flow-chart we present in this paper offers a solution for practitioners to reduce the electrical consumption while a technological processing of special materials is necessary. The computer program and the database can be easily adapted to any technological processes (conventional or not).

## 1. General presentation of a non-conventional technology

The complex electric erosion (CEE) is one non-conventional technological processing characterized by the overlapping in space in time of the two classic processing procedures: the electric erosion and electro-chemical erosion, with mechanic depassivation ([HH99]) and it consist of the following physic-chemic, mechanic and electric process:
- because of the processed object (PO) and of the transfer object (TO) connection to a source of continuous current, an electric field (E)

appears between the two electrodes, leading to the emergence of a substance transportation (due to the chemical reactions) from the liquid working environment (WE) in which the processing takes place and on the PO surface;
- when the film laid down on the TO surface gets a certain thickness, the PO chemical dissolving process stops;
- because of the PO dislocation towards TO and of the contact pressure between the electrodes, the film is removed and the two pieces come in contact;
- the electrical discharges in impulse occur, melting and vaporizing the PO surface;
- the newly appeared craters on the PO surface permit the restarting of the processing process.

Among the electrical, mechanical, environmental factors influencing the results of the complex electrical erosion processing, the working current intensity/the current density/the induced power in the working space and the pressure between the two electrodes have the biggest influence.

At first sight, maximizing productivity could be done by increasing the introduced power in the working space (WS). Yet, experimentally in [Her95] and [Kar04], there has been determined that the increasing of the tension over 30V and of the current over 100 A leads to the instability of the process and to bad results (productivity, the quality of the surface) - the phenomenon is explained by the existence of a certain current density, which, when surpassed, makes the electrical discharges between electrodes turn from the impulse phase to the one of an electrical arch, hard to control, with huge material extractions from the PO surface and rapid TO degradation.

## 2. Non-conventional debiting. The mathematical modeling

Our study presents the use of an interactive informational system which prepares the processing thorough an unconventional method (CEE), selects the experimental data, models and simulates the developing processes, optimizes and permits the running of the CEE process [Kar04].

The modeling of the debiting process, using an ordinary or a non-conventional processing technique can be implemented by mathematical instruments (as numeric calculus), software instruments (numerical software), and methods of artificial intelligence.

For the modeling of the dependency of the processing time ($t_p$) on the induced power (P), we tried polynomial functions as presented in [O+99], regarding to the most usual dependencies presented in the literature ([Kar04]):

$$t_{p1} = a_0 + a_1 \cdot P \ [s] \tag{1.1}$$
$$t_{p2} = a_0 + a_1 \cdot P + a_2 \cdot P^2 \ [s] \tag{1.2}$$
$$t_{p3} = a_0 + a_1 \cdot P + a_2 \cdot P^2 + a_3 \cdot P^3 \ [s] \tag{1.3}$$

In order to determine the mathematic pattern of the processing time dependency to the PC52 carbon steel debiting, there was experimented ([O+99]) the debiting of some samples on a CEE processing machine, using different electrical regimes for current voltage and intensity and also for the impedance parameters.

These experiments used PC52 samples (processed object, PO) with 20 mm diameter, debited on a MEC-50 debiting machine in the Non-Conventional Technologies Laboratory of the Mechanical Technologies Department, "Politehnica" University of Timişoara, Romania. The cutting tools (transfer object, TO) are made by OL37 with 200 mm diameter and 1 mm thickness.

The results of the experiments are displayed in Table 1 and graphic represented in Figure 1, obtained using MathCAD application (MathSoft Inc. Cambridge, MA, USA).

To determinate the coefficients in the (1.1) – (1.3) functions we used a polynomial interpolation by less square method, using a computer software developed for this situation.

The following results were obtained (Figure 2):

$$t_{p1} = 139.8528 - 0.0227 \cdot P \ [s] \tag{1.4}$$
$$t_{p2} = 173.1836 - 0.0664 \cdot P + 6.3902 \cdot 10^{-6} \cdot P^2 \ [s] \tag{1.5}$$
$$t_{p3} = 203.1861 - 0.1286 \cdot P + 2.9231 \cdot 10^{-5} \cdot P^2 - 2.1012 \cdot 10^{-9} \cdot P^3 \ [s] \tag{1.6}$$

The validation of the mathematical model was obtained by comparing the experimental results with the results obtained by the application of the polynomial function, for input values (induced power) others than the experimental ones. We calculated the deviation in each cases and the correct model appears for the minimum dispersion.

*Table 1: Experimental results for TO=OL37*

| Regime | U [V] | I [A] | P [W] | $t_p$ [s] |
|---|---|---|---|---|
| I | 16 | 30 | 480 | 152 |
|   | 18 | 35 | 630 | 140 |
|   | 20 | 40 | 800 | 128 |
| II | 14 | 25 | 350 | 155 |
|    | 15.5 | 35 | 542.5 | 135 |
|    | 20 | 50 | 1000 | 112 |
| III | 20 | 80 | 1600 | 52 |
|     | 25 | 120 | 3000 | 45 |
|     | 35 | 200 | 7000 | 36 |
| IV | 25 | 120 | 3000 | 30 |
|    | 30 | 150 | 4500 | 23 |
|    | 35 | 200 | 7000 | 15 |

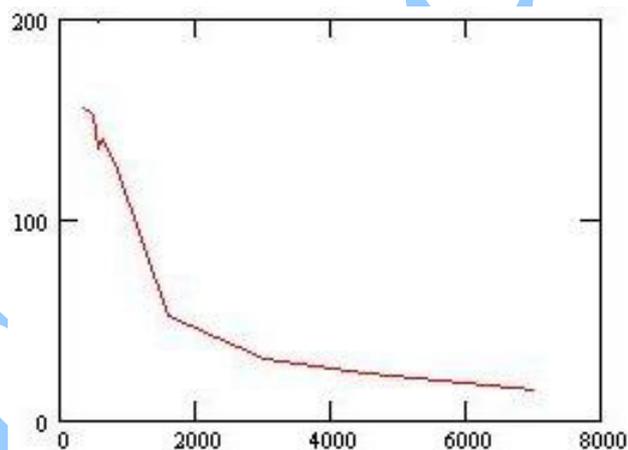

*Figure 1: The processing time $t_p$ dependency on the induced power P*

These results are presented in Table 2 and the simultaneous representation of the modeling polynomial function is depicted in Figure 3.

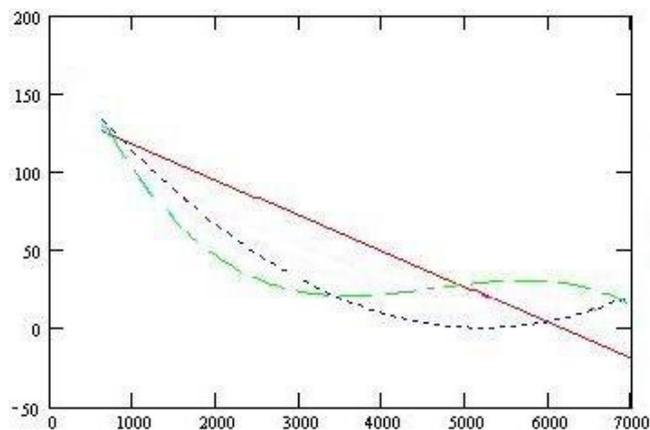

*Figure 2: The $t_p=f(P)$ dependency using modeling functions (red=(1.4), blue=(1.5), green=(1.6))*

*Table 2: The standard deviation for the $t_p=f(P)$ dependency*

| Function | Deviation | Notes |
|---|---|---|
| (1.4) | 97.228 | |
| (1.5) | 37.339 | |
| (1.6) | 11.295 | Optimum |

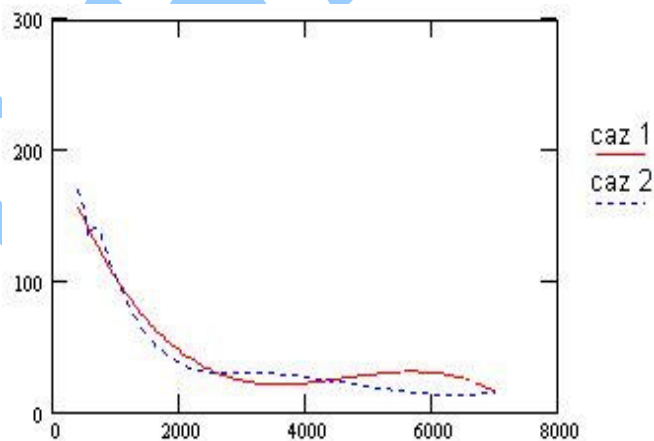

*Figure 3: The $t_p=f(P)$ dependency using modeling functions (case 1=experimental results, case 2=modeling function)*

193

### 3. Non-conventional debiting. The simulation

The simulation can be often used, in technological processes, to establish the optimum conditions for the processing parameters (in this paper, to minimize the processing time at the debiting using the CEE).

But to establish the optimum only as result of a mathematical pattern and simulation must be applied in conjunction with the input variables made by the economic and technological evolution of the end-user, by the material possibilities and resources, by the necessities and the acknowledge level.

Mathematically speaking, the optimization consists on the determination of the input values that maximizes or minimizes the results. Practically, it is demonstrated that a simultaneously optimization of all outputs cannot be obtained, due to the extreme interdependency between all parameters. As higher we can push a result values, as lower can be inhered another result ([Kar04]). That's the reason why the user must select the basic (the main) indicator and the set of other, less important, indicators.

The main purpose of the simulation is to establish the optimum working regime to debit the sample of PO.

The answer to this problem starts from the mathematical functions $t_p=f(P)$ and continues computing the minimum of these functions. This calculus can be made by three ways as introduced in [O+99]:
- by derivation the mathematical function and solving the new one;
- by looking to all admissible input values and searching the minimum value;
- by computing the $P=f^{-1}(t_p)$ functions.

The simulation consists on the following algorithm ([O+99]):
- first, the obtaining of the mathematical model of the process;
- then, the model testing;
- last, the use of the model.

There are two main methods to obtain the optimum of a mathematical function using the numeric simulation ([N+04]):
- the method of pure random exploration, using a set of points monotonous distributed into a domain and memorizing the optimum value after each loop;
- the method of controlled random exploration, using a set of mathematical and comparative operations into each loop; this method us longer but reduces the number of loops.

We used the second simulation method to obtain the data presented in this paper and these two algorithms above can be used to calculate the optimum values of a single- or two-variable polynomial function.

A numeric algorithm offers a shorter way to obtain the minimum value of the debiting time ($t_p$). The reducing of the admissible range method can provide the optimum as presented in Table 3.

*Table 3: The optimum values*

| Function | Optimum P | Optimum $t_p$ |
|---|---|---|
| (1.1) | 3000 | 71.75 |
| (1.2) | 2800 | 65.55 |
| (1.3) | 2750 | 62.04 |

## 4. Non-conventional debiting. The informational system

If these methods are implemented simultaneously with a database containing the results of previous experiments, with a software application that computes the mathematical model of the process, as presented in [HH99] and, finally, with an interface computer-processing machine to manage the debiting operation as proposed in [HH99], the economical results of the non-conventional procedures can be more and more interesting and easier to implement.

Nowadays, data bases are preferentially used for the management of large quantity of information. These have been minutely described in studies and by calculi done in a short time. Regarding the processing process through unconventional technologies, it has been determined that the best choice would be a relational data base containing the tables presented in [HH99], [Kar04] and [N+03].

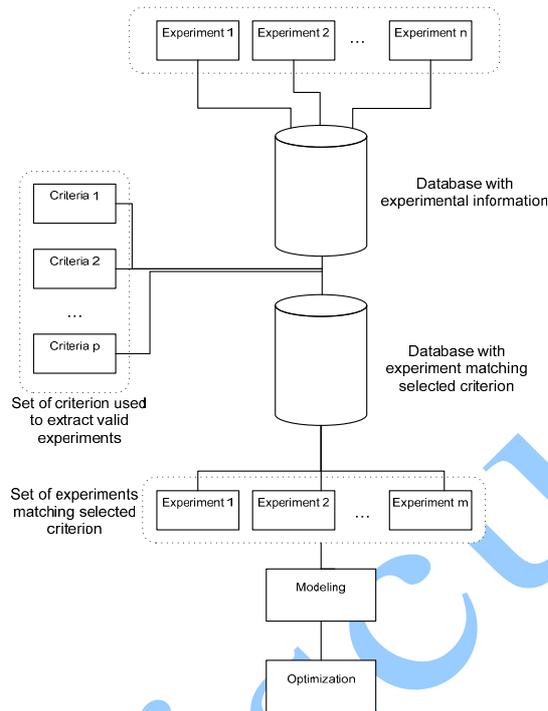

*Figure 4: Informational system for non-conventional debiting*

Using the experimental information contained by this data base, the following operations can be done:
- The creation and the loading of the data base with experimental information;
- Organizing the data base according to the desired criterion;
- Obtaining the lists with the performed experiments, arranged according to the user's wish:
    - according to a certain material of the PO and TO;
    - according to a processing machine;
    - according to a certain processing operation.
- Determining all the experiments about the processing of a certain material;
- Calculating the optimal processing conditions for a certain PO, according to the existent information from the data base, through mono and multivariable modeling, linear or non-linear optimizing.

All those steps are graphically pointed out in Figure 4.

**Conclusions**

Economically, the complex electric erosion is to be preferred to other traditional methods there where the processed object (PO) is made of a material very difficult to process: special steels, diverse stainless alloys etc.

By using a data base regarding the results of the performed experiments, with the previously presented structure, containing 1300 pieces of information about 50 materials which can be processed by electro-chemical complex erosion (alloyed steels, stainless steels, carburets, cast iron etc.), completed with the elaboration of automatic algorithm to determine the debiting optimal conditions of a material, according to the known physical-chemic and mechanic feature, the debiting process through this unconventional technological method can bring a plus of efficiency which is so necessary in today's conditions of permanent competition.

**References**

[Her95]    Herman, R. (1995), *Contribuţii la optimizarea realizării fantelor prin eroziune electrică complexă*, Ph.D. Thesis, Technical University of Timişoara

[HH99]     Herman, R. and Herman, M. (1999), "Consideraţii privind un sistem generalizat de acţiune tehnologică în procesul de prelucrarea prin eroziune electrică complexă"*, Tehnologii neconvenţionale* 2/1999, Augusta, Timişoara

[KN01]     Karnyanszky, T.M. and Nanu, A. (2001), "A Study about the Complex Electric Erosion Processing Modelling", *Revista de tehnologii neconvenţionale* 2/2001, Augusta, Timişoara, 7-10

[Kar04]    Karnyanszky, T.M. (2004), *Contribuţii la conducerea automată a prelucrării dimensionale prin eroziune electrică complexă*, Ph.D. Thesis, "Politehnica" University of Timişoara

[N+03]     Nanu, A. et al. (2003), *Tratat de tehnologii neconvenţionale. Vol. I – Tehnologiile neconvenţionale la început de mileniu*, Asociaţia Română pentru Tehnologii Neconvenţionale,

Academia Română, Academia de Ştiinţe Tehnice din România, Augusta, Timişoara

[N+04]  Nanu, A. et al. (2004), *Tratat de tehnologii neconvenţionale. Vol. V – Prelucrarea prin eroziune complexă electrică-electrochimică*, Asociaţia Română pentru Tehnologii Neconvenţionale, Academia Română, Academia de Ştiinţe Tehnice din România, Augusta, Timişoara

[O+99]  Oprean, C., Nanu, D., Ţîţu, M., Cicală, E. and Vannes A.B. (1999), "Software universal pentru modelare, optimizare şi conducere asistată a proceselor tehnologice", *Revista de tehnologii neconvenţionale*, 1/1999, Augusta, Timişoara, 107-114